\documentclass[useAMS,usenatbib]{mn2e}
\usepackage{graphicx}
\usepackage{color}
\usepackage{ulem}
\def\apj{ApJ}
\def\apjs{ApJS}

\def\aap{A\&A}
\def\aaps{A\&AS}

\def\mnras{MNRAS}

\def\pasp{PASP}
\def\nat{Nature}

\def\araa{Ann.Rev.Astron.Astrophys.}

\title[J2229$+$2725, an extremely low-metallicity galaxy]
{J2229$+$2725: an extremely low-metallicity dwarf compact 
star-forming galaxy with an exceptionally high 
[O~{\sc iii}]$\lambda$5007/[O~{\sc ii}]$\lambda$3727 flux ratio of 53}
\author[Y. I. Izotov et al.]{Y. I.\ Izotov$^{1}$\thanks{Corresponding author: yizotov@bitp.kiev.ua},
T. X.\ Thuan$^{2}$ and N. G.\ Guseva$^{1}$\\
                $^{1}$Bogolyubov Institute for Theoretical Physics,
                     National Academy of Sciences of Ukraine,
                     14-b Metrolohichna str., Kyiv, 03143, Ukraine,\\
                $^{2}$Astronomy Department, University of Virginia, 
                     P.O. Box 400325, Charlottesville, VA 22904-4325,\\
}
\begin{document}


\pagerange{\pageref{firstpage}--\pageref{lastpage}} \pubyear{2012}

\maketitle

\label{firstpage}

\begin{abstract}
Using the Large Binocular 
Telescope (LBT)/Multi-Object Dual Spectrograph (MODS), we have obtained optical spectroscopy of one of the most 
metal-poor dwarf star-forming galaxies (SFG) in the local Universe, 
J2229$+$2725,. 
This galaxy with a 
redshift $z$=0.0762 was selected from the
Data Release 16 (DR16) of the Sloan Digital Sky Survey (SDSS). 
Its properties derived from the LBT observations are most extreme among 
SFGs in several ways. Its oxygen abundance   
12~+~logO/H = 7.085$\pm$0.031 is among the lowest ever observed for a SFG. With its very low metallicity, 
an absolute magnitude $M_g$ = $-$16.39 mag, a low stellar mass 
$M_\star$ = 9.1$\times$10$^6$~M$_\odot$ and a very low mass-to-light ratio 
$M_\star$/$L_g$~$\sim$~0.0166 (in solar units), J2229$+$2725 deviates strongly 
from the luminosity-metallicity relation defined by the bulk of 
the SFGs in the SDSS. J2229$+$2725 has a very high  specific star-formation rate
sSFR $\sim$ 75 Gyr$^{-1}$, indicating very active ongoing star formation. 
Three other features of J2229$+$2725 are most striking, being the most extreme 
among lowest-metallicity SFGs: 1) a ratio O$_{32}$ = 
$I$([O~{\sc iii}]$\lambda$5007)/$I$([O~{\sc ii}]$\lambda$3727) $\sim$ 53, 
2) an equivalent width of the 
H$\beta$ emission line EW(H$\beta$) of 577\AA, and 3) an electron number
density of $\sim$ 1000 cm$^{-3}$. These properties imply that 
the starburst in J2229$+$2725 is very young.
Using the extremely high O$_{32}$ in J2229$+$2725, we have improved the 
strong-line calibration for the 
determination of oxygen abundances in the most metal-deficient galaxies, 
in the range 12 + logO/H $\la$ 7.3.
\end{abstract}


\section{Introduction}\label{sec:INT}

Extremely metal-deficient (XMD) low-redshift star-forming galaxies (SFGs), with 
oxygen abundances as low as 12~+~logO/H $\sim$ 6.9 -- 7.1
have been recently uncovered in a number of studies. Thus, \citet{H16} have 
reported 12~+~logO/H = 7.02$\pm$0.03 in the galaxy AGC~198691. \citet{H17}
have found the oxygen abundance in the galaxy Little Cub 
to be 12~+~logO/H = 7.13$\pm$0.08.
\citet{I18a} and \citet{Iz19a}, based on Large Binocular Telescope (LBT)
observations, derived oxygen abundances in J0811$+$4730 and J1234$+$3901
of 6.98$\pm$0.02 and 7.035$\pm$0.026, respectively. 
Both these galaxies have been selected from the Sloan Digital Sky Survey 
(SDSS). \citet{Ko20} have recently reported the discovery of the galaxy 
J1631$+$4426 with 12~+~logO/H = 6.90$\pm$0.03. In all these studies, the 
oxygen abundances have been derived by the direct $T_{\rm e}$ method from
spectra of the entire galaxy. The common
characteristics of the galaxies mentioned above are their low stellar mass
$M_\star$ $\la$ 10$^7$ M$_\odot$ and a very compact structure. 
Besides these galaxies, very low oxygen abundances of 7.01$\pm$0.07, 
6.98$\pm$0.06, 
6.86$\pm$0.14 have also been found by \citet{I09} in three individual H~{\sc ii}
regions of the XMD SBS~0335$-$052W. The value of 
6.96$\pm$0.09 has been obtained by \citet{An19} in one of the H~{\sc ii} 
regions in the dwarf irregular galaxy DDO~68. However, some other regions in 
SBS~0335$-$052W and DDO~68 show oxygen abundances above 7.1.

The importance of finding and studying these galaxies is emphasized by
the fact that they share many of the same properties with the dwarf galaxies at 
high redshifts and thus may be considered as the their best local counterparts. 
Therefore, the low-redshift XMD SFGs with 12~+~logO/H~$\sim$~7.0 are excellent 
laboratories for studying the physical
conditions in dwarf galaxies at redshifts $z$~$\sim$~5--10, during the epoch of the
Universe's reionisation.

\begin{figure}
\centering
\includegraphics[angle=0,width=0.7\linewidth]{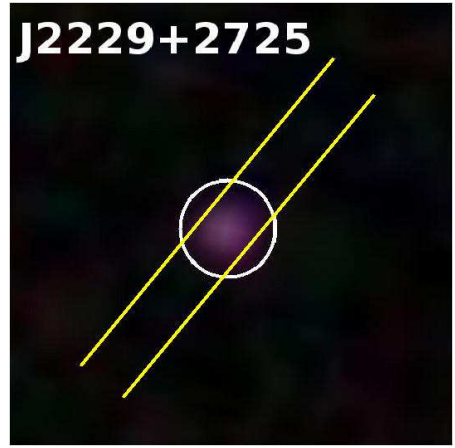}
\caption{12 arcsec $\times$ 12 arcsec region with the color composite SDSS image
of J2229+2725. The 2 arcsec SDSS spectroscopic aperture and 1.2 arcsec wide 
LBT/MODS slit are indicated by a white circle and yellow parallel lines, 
respectively.
}
\label{fig1}
\end{figure}

In this paper, we present LBT\footnote{The LBT 
is an international collaboration among institutions in the United States, 
Italy and Germany. LBT Corporation partners are: The University of Arizona on 
behalf of the Arizona university system; Istituto Nazionale di Astrofisica, 
Italy; LBT Beteiligungsgesellschaft, Germany, representing the Max-Planck 
Society, 
the Astrophysical Institute Potsdam, and Heidelberg University; The Ohio State 
University, and The Research Corporation, on behalf of The University of Notre 
Dame, University of Minnesota and University of Virginia.} spectroscopic
observations of a new SFG, J2229$+$2725. 
The galaxy was selected from the SDSS Data Release 16 (DR16) data base 
\citep{Ah20} as a candidate to be an extremely low-metallicity object, using 
criteria such as the simultaneous presence of a low [O~{\sc iii}]$\lambda$5007/H$\beta$ ratio 
of $\la$ 3 and of a weak or undetected [N~{\sc ii}]$\lambda$6584 emission line
with a flux $\la$ 1/100 that of the H$\alpha$ emission line 
\citep[e.g. ][]{Iz19b}. We also selected J2229$+$2725 because of its extremely 
high O$_{32}$ value, defined as the ratio
$I$([O~{\sc iii}]$\lambda$5007)/$I$([O~{\sc ii}]$\lambda$3727). This high ratio 
implies that the galaxy may contain density-bounded H~{\sc ii} regions, and 
thus be a potential Lyman continuum leaker (e.g. Jaskot \& Oey 2013) 
and a good local counterpart to the high-redshift dwarf galaxies thought to be 
responsible for the reionisation of the early Universe 
\citep*{O09,WC09,Y11,M13,B15,I16a,I16b,I18b,I18c}. Its coordinates, redshift 
and other characteristics
obtained from the photometric and spectroscopic SDSS, the {\sl Galaxy Evolution 
Explorer} ({\sl GALEX}) and LBT data are presented in Table \ref{tab1}.
The SDSS image of J2229+2725 is displayed in Fig.~\ref{fig1}. For comparison,
we show the round 2 arcsec SDSS spectroscopic aperture centered on the galaxy and 
the 1.2 arcsec wide LBT slit. The
galaxy is very compact, with a full width at half maximum (FWHM) of 0.9 
arcsec, comparable to the median Apache Point Observatory (APO) and 
LBT Observatory (LBTO) seeings during the observations and corresponding to a linear size of 
1.3 kpc. The actual size of J2229+2725 is likely smaller and comparable to the sub-kpc sizes
of compact star-forming galaxies at similar redshifts with O$_{32}$ = 22 -- 39, observed with the {\sl Hubble Space Telescope} ({\sl HST}) 
\citep{Iz20}. No substantial emission from J2229+2725 is seen outside the SDSS spectroscopic 
aperture and the LBT slit in Fig.~\ref{fig1}, implying that aperture corrections are small.

\begin{table}
\caption{Observed and derived characteristics of J2229$+$2725 \label{tab1}}
\begin{tabular}{lr} \hline
Parameter                 &  J2229$+$2725       \\ \hline
R.A.(J2000)               &  22:29:33.19 \\
Dec.(J2000)               & +27:25:25.60 \\
  $z$                     &  0.07622     \\
{\sl GALEX}  FUV, mag     &   21.45$\pm$0.28      \\
{\sl GALEX}  NUV, mag     &   21.88$\pm$0.37      \\
SDSS $u$, mag             &   21.96$\pm$0.15      \\
SDSS $g$, mag             &   21.47$\pm$0.04      \\
SDSS $r$, mag             &   22.27$\pm$0.11      \\
SDSS $i$, mag             &   20.97$\pm$0.05      \\
SDSS $z$, mag             &   21.43$\pm$0.33      \\
   $D_L$, Mpc$^{*}$        &       347     \\
 $M_g$, mag$^\dag$         & $-$16.39$\pm$0.06     \\
log $L_g$/L$_{g,\odot}$$^\ddag$&     8.74     \\
log $M_\star$/M$_\odot$$^{\dag\dag}$&   6.96       \\
$M_\star$/$L_g$, M$_\odot$/L$_{g,\odot}$& 0.0166       \\
$L$(H$\beta$), erg s$^{-1}$$^{**}$&(3.1$\pm$0.3)$\times$10$^{40}$\\
SFR, M$_\odot$yr$^{-1}$$^{\ddag\ddag}$  &     0.68$\pm$0.06 \\
sSFR, Gyr$^{-1}$          &     75 \\
  12+logO/H$^{\dag\dag\dag}$              &7.085$\pm$0.031 \\
\hline
  \end{tabular}


\noindent$^{*}$Luminosity distance.

\noindent$^\dag$Absolute magnitude corrected for Milky Way extinction.

\noindent$^\ddag$log of the SDSS $g$-band luminosity corrected for Milky Way extinction.

\noindent$^{\dag\dag}$Stellar mass derived from the extinction-corrected SDSS spectrum.

\noindent$^{**}$H$\beta$ luminosity derived from the extinction-corrected SDSS 
spectrum.

\noindent$^{\ddag\ddag}$Star formation rate derived from the \citet{K98} relation
 using the extinction-corrected H$\beta$ luminosity.

\noindent$^{\dag\dag\dag}$Oxygen abundance derived from the LBT spectrum.

  \end{table}

\begin{figure*}[t]
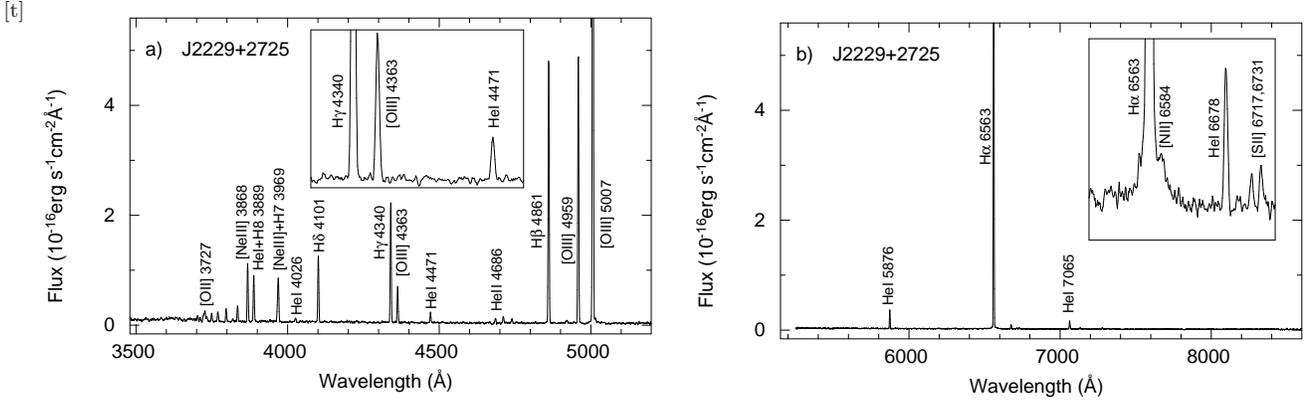

\hbox{
\includegraphics[angle=-90,width=0.46\linewidth]{f2229+2725b_1.ps}
\hspace{0.4cm}\includegraphics[angle=-90,width=0.46\linewidth]{f2229+2725r_1.ps}
}
\caption{The rest-frame LBT spectrum of J2229$+$2725.
Insets in {\bf a)} and {\bf b)} show expanded parts of the spectral regions around 
the H$\gamma$ and H$\alpha$ emission lines, respectively, for a better view of 
weak features. Some emission lines are labelled. 
}
\label{fig2}
\end{figure*}

\section{Observations and data reduction}\label{sec:observations}

We have obtained LBT long-slit spectrophotometric observations of J2229$+$2725 
on 15 September, 2020 in the twin binocular mode, using the MODS1 and MODS2
spectrographs\footnote{This paper used data obtained with the MODS 
spectrographs built with
funding from NSF grant AST-9987045 and the NSF Telescope System
Instrumentation Program (TSIP), with additional funds from the Ohio
Board of Regents and the Ohio State University Office of Research.}.
The night was nearly clear with some small clouds moving through.
Spectra were 
obtained in the wavelength range 3200 -- 10000\AA\ with a 1.2 arcsec wide slit, 
resulting in a resolving power $R$ $\sim$ 2000.
The seeing during the observations was 1.0  arcsec.

Four 900~s subexposures were obtained in both the blue and red ranges separately
with MODS1 and MODS2, resulting in a total exposure time of 2$\times$3600~s, 
counting both spectrographs. The airmass during observations was small, equal to 1.01.
Thus, the effect of atmospheric refraction is small 
for all subexposures \citep[see ][]{F82}. 

The spectrum of the spectrophotometric standard star 
BD+33~2642 was obtained during the same night, with a 5 arcsec wide slit 
for flux calibration and correction for telluric absorption in the red part.
Additionally, calibration frames of biases, flats and comparison lamps 
were obtained during the same period with the same setups of MODS1 and MODS2.

Bias subtraction, flat field correction, wavelength and flux calibration
were done with the MODS Basic CCD Reduction package 
{\sc modsccdred}\footnote{http://www.astronomy.ohio-state.edu/MODS/Manuals/ MODSCCDRed.pdf} and 
{\sc iraf}\footnote{{\sc iraf} is distributed by the 
National Optical Astronomy Observatories, which are operated by the Association
of Universities for Research in Astronomy, Inc., under cooperative agreement 
with the National Science Foundation.}. After these reduction 
steps, MODS1 and MODS2 subexposures were co-added and one-dimensional 
spectra of J2229$+$2725 in the blue and red ranges were extracted in a 1.2 
arcsec aperture along the spatial axis, tracing pixels with the highest fluxes
along the wavelength axis, using the {\sc iraf apall} routine. 
This aperture includes nearly 
all emission from J2229+2725 (Fig~\ref{fig1}). This is the case because the observed LBT H$\beta$
flux of 2.145$\times$10$^{-15}$ erg s$^{-1}$cm$^{-2}$ compares well to the 
H$\beta$ flux of 1.77$\times$10$^{-15}$ erg s$^{-1}$cm$^{-2}$ measured in the
SDSS spectrum. The LBT spectra exhibit intense emission 
lines, including a strong [O~{\sc iii}]$\lambda$4363 emission line. They are 
shown in Fig. \ref{fig2}. We note that the spatial extent of the galaxy 
along the slit is small, with a FWHM of 0.7 arcsec and a width of 1.5 arcsec at 
the 10 per cent level of maximal intensity. These angular sizes correspond to 
linear sizes of 1.0 and 2.2 kpc, respectively.

The observed emission-line fluxes and their errors were measured using the 
{\sc iraf splot} routine. They were corrected iteratively for extinction and underlying 
stellar absorption, derived from the observed decrement 
of the hydrogen Balmer emission lines H$\alpha$, H$\beta$, H$\gamma$, 
H$\delta$, H9, H10, H11, H12, as described in \citet{ITL94}. 
The equivalent widths of the underlying stellar Balmer 
absorption lines are assumed to be the same for each line. 
The extinction-corrected fluxes together with the extinction 
coefficient $C$(H$\beta$), the observed H$\beta$ emission-line flux 
$F$(H$\beta$), the rest-frame equivalent width EW(H$\beta$) of the H$\beta$ 
emission line, and the equivalent width of the Balmer absorption lines
are shown in Table~\ref{tab2}. The fluxes of all hydrogen emission lines 
relative to H$\beta$ corrected for extinction and underlying 
absorption are close to the theoretical recombination values \citep{SH95}.

\begin{figure}
\centering{
\includegraphics[angle=-90,width=0.90\linewidth]{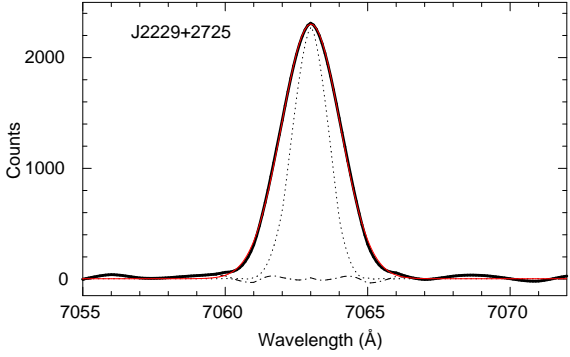}
}
\caption{The H$\alpha$ profile in the medium-resolution DIS spectrum of 
J2229$+$2725. The observed spectrum is shown by a black solid line while the Gaussian 
fit is represented by a red solid line. The black dash-dotted line shows the
residual spectrum after subtraction of the H$\alpha$ fit. The instrumental
profile is represented by a black dotted line.}
\label{fig3}
\end{figure}

A medium-resolution spectrum of J2229$+$2725 with the H$\alpha$ emission line 
was obtained on 20 September, 2020 with the Double Imaging Spectrograph (DIS) 
mounted on the 3.5 meter Apache Point Observatory (APO) telescope\footnote{Based
on observations with the the 3.5m Apache Point Observatory (APO). The Apache 
Point Observatory 3.5-meter telescope is owned and operated by the 
Astrophysical Research Consortium.}. We use the R1200 grating with a linear 
dispersion 0.58 \AA/pix and a resolving power of $\sim$~6000. The above instrumental set-up gave a spatial scale 
along the slit of 0.4 arcsec pixel$^{-1}$ and a spectral coverage of
$\sim$~1160\AA\ over 2000 pixels. A spectroscopic standard star G191b2b
was observed for flux calibration. Spectra of He-Ne-Ar comparison arcs were 
obtained at the end of the night for wavelength calibration. The part of the
DIS spectrum with the H$\alpha$ emission line is shown in Fig.~\ref{fig3}.

\begin{figure}
\centering{
\includegraphics[angle=-90,width=0.90\linewidth]{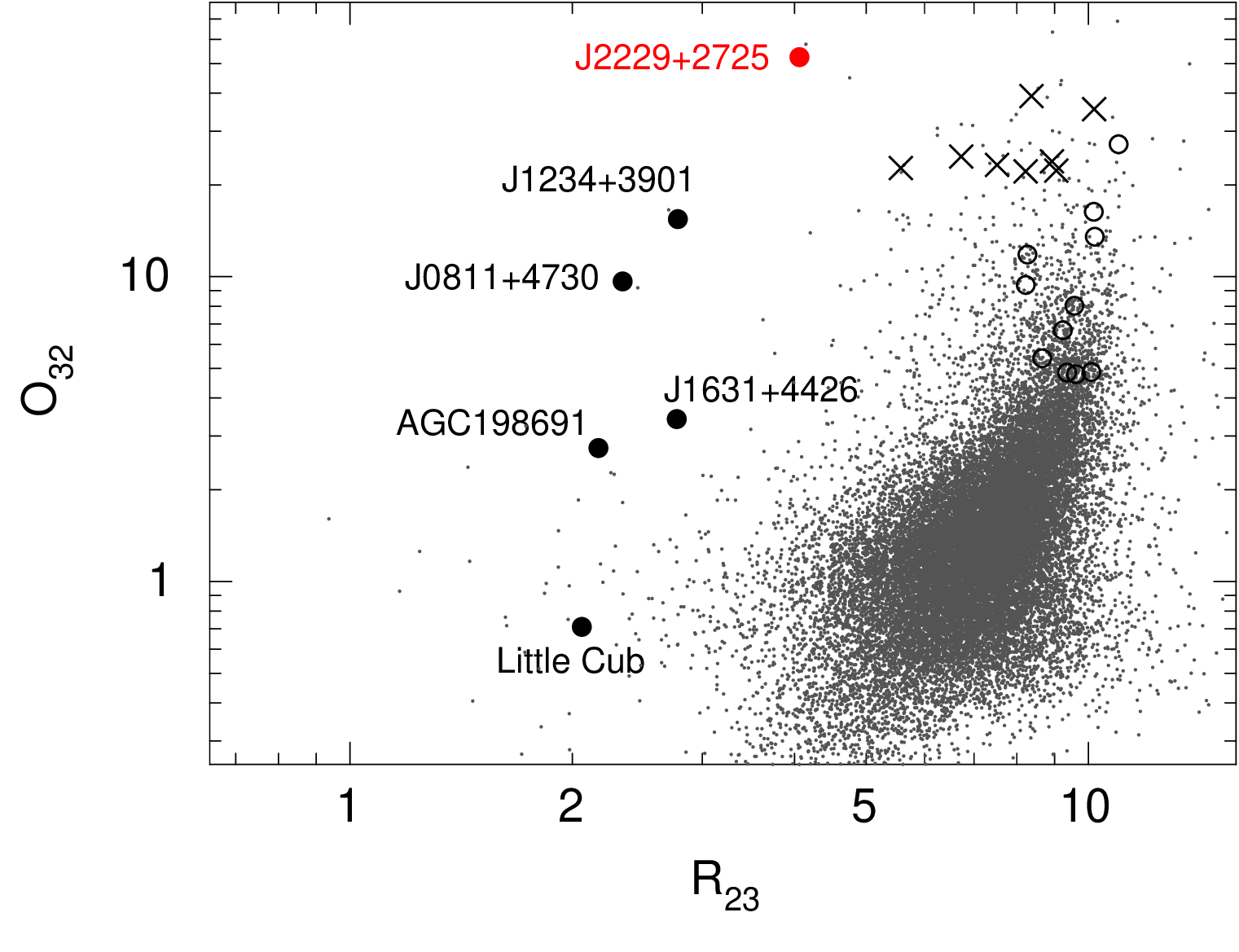}
}
\caption{The O$_{32}$ -- R$_{23}$ diagram for compact SFGs, where O$_{32}$= 
$I$([O~{\sc iii}]$\lambda$5007)/$I$([O~{\sc ii}]$\lambda$3727) and R$_{23}$ = 
$I$([O~{\sc ii}]$\lambda$3727 + [O~{\sc iii}]$\lambda$4959 + 
[O~{\sc iii}]$\lambda$5007)/$I$(H$\beta$). The lowest-metallicity SFGs known
with 12 + log O/H $<$ 7.1
from \citet[][A198691]{H16}, \citet[][Little Cub]{H17}, 
\citet[][J0811$+$4730]{I18a}, \citet[][J1234$+$3901]{Iz19a}, 
\citet[][J1631$+$4426]{Ko20} and this paper (J2229$+$2725) are shown 
by labelled filled circles.
LyC leakers from 
\citet{I16a,I16b,I18b,I18c} are shown by open circles.  SFGs with the highest
O$_{32}$ from \citet{Iz20} are represented by crosses and compact SFGs from the 
SDSS DR16 by grey dots.}
\label{fig4}
\end{figure}

We note two features in the LBT spectrum of J2229$+$2725, which have never been seen before 
in the spectra of other SFGs, including those with the lowest heavy element 
abundances. First, the O$_{32}$ ratio of 53 is enormous compared to that in 
other SFGs. It is comparable to the value of 57 derived from the extinction-corrected
SDSS spectrum. We note however that the [O~{\sc iii}] $\lambda$5007 emission
is clipped in the SDSS spectrum. Therefore, in the calculation of O$_{32}$, we 
have adopted 
$F$([O~{\sc iii}] $\lambda$5007) = 3 $\times$ $F$([O~{\sc iii}] $\lambda$4959).
The high value of O$_{32}$ is better demonstrated in Fig.~\ref{fig4}, where 
we show the diagram O$_{32}$ -- R$_{23}$, with R$_{23}$ = 
$I$([O~{\sc ii}]$\lambda$3727 + [O~{\sc iii}]$\lambda$4959 + [O~{\sc iii}]$\lambda$5007)/$I$(H$\beta$), for the lowest-metallicity compact
SFGs known (with spectra obtained for the entire galaxy), compact SFGs from the 
SDSS and confirmed low-$z$ LyC leaking galaxies.
The value O$_{32}$ for J2229$+$2725 is far above the values for other 
SFGs with the lowest oxygen abundances known, whereas it is offset to low R$_{23}$
compared to the main sequence defined by the SDSS SFGs, due to its extremely low
oxygen abundance. Second, the 
H$\beta$ equivalent width EW(H$\beta$) of $\sim$~577\AA\ is also enormous.
Both features indicate that the starburst in J2229$+$2725 is very 
young, with an age $\la$~1~--~2~Myr.
Its spectrum shows a relatively strong He~{\sc ii} $\lambda$4686 emission 
line, with a flux $\sim$ 2.1 per cent that of the H$\beta$ emission line, 
similar to that in integrated spectra of other lowest-metallicity
compact star-forming galaxies 
\citep[e.g. in J0811+4730 and J1234+3901,][]{I18a,Iz19a} and indicative of the 
presence of hard ionising radiation.

\begin{table}
\caption{Extinction-corrected emission-line flux ratios$^{*}$ \label{tab2}}
\begin{tabular}{lr} \hline
Line& \multicolumn{1}{c}{J2229$+$2725}   \\ \hline
3721.94 H14                     &  1.43$\pm$0.21\\
3727.00 [O {\sc ii}]            &  5.72$\pm$0.39\\
3734.37 H13                     &  2.63$\pm$0.38\\
3750.15 H12                     &  4.45$\pm$0.59\\
3770.63 H11                     &  5.22$\pm$0.61\\
3797.90 H10                     &  6.36$\pm$0.54\\
3819.64 He {\sc i}              &  1.52$\pm$0.24\\
3835.39 H9                      &  7.35$\pm$0.54\\
3868.76 [Ne {\sc iii}]          & 22.49$\pm$0.81\\
3889.00 He {\sc i}+H8           & 19.97$\pm$0.84\\
3968.00 [Ne {\sc iii}]+H7       & 23.81$\pm$0.92\\
4026.19 He {\sc i}              &  1.95$\pm$0.32\\
4101.74 H$\delta$               & 26.34$\pm$0.97\\
4227.20 [Fe {\sc v}]            &  1.35$\pm$0.21\\
4340.47 H$\gamma$               & 46.57$\pm$1.48\\
4363.21 [O {\sc iii}]           & 13.96$\pm$0.56\\
4471.48 He {\sc i}              &  3.92$\pm$0.28\\
4685.94 He {\sc ii}             &  2.12$\pm$0.24\\
4712.00 [Ar {\sc iv}]+He {\sc i}&  2.85$\pm$0.25\\
4740.20 [Ar {\sc iv}]           &  1.89$\pm$0.22\\
4861.33 H$\beta$                &100.00$\pm$2.97\\
4921.93 He {\sc i}              &  1.60$\pm$0.21\\
4958.92 [O {\sc iii}]           &100.01$\pm$2.96\\
5006.80 [O {\sc iii}]           &300.36$\pm$8.79\\
5875.60 He {\sc i}              &  9.47$\pm$0.31\\
6312.10 [S~{\sc ii}]            &  0.36$\pm$0.12\\
6562.80 H$\alpha$               &269.65$\pm$8.50\\
6583.40 [N~{\sc ii}]            &  0.26$\pm$0.12\\
6678.10 He {\sc i}              &  2.31$\pm$0.12\\
6716.40 [S~{\sc ii}]             &  0.56$\pm$0.06\\
6730.80 [S~{\sc ii}]             &  0.62$\pm$0.06\\
7065.30 He {\sc i}              &  4.51$\pm$0.27\\
7135.80 [Ar~{\sc iii}]           &  0.79$\pm$0.08\\
7281.35 He {\sc i}              &  0.79$\pm$0.06\\ \\
$C$(H$\beta$)$^{\dag}$         &\multicolumn{1}{c}{0.060$\pm$0.038}\\
$F$(H$\beta$)$^{\ddag}$        &\multicolumn{1}{c}{21.45$\pm$0.12}\\
EW(H$\beta$)$^{**}$           &\multicolumn{1}{c}{577.2$\pm$3.6}\\
EW(abs)$^{**}$                &\multicolumn{1}{c}{3.4$\pm$1.1}\\
\hline
  \end{tabular}

\hbox{$^{*}$in units 100$\times$$I(\lambda)$/$I$(H$\beta$).} 

\hbox{$^{\dag}$Extinction coefficient, derived from the observed hydrogen} 

\hbox{\,~Balmer decrement.}

\hbox{$^{\ddag}$Observed flux in units of 10$^{-16}$ erg s$^{-1}$ cm$^{-2}$.}

\hbox{$^{**}$Equivalent width in \AA.}

  \end{table}

\begin{table}
\caption{Electron temperatures, electron number density 
and heavy element abundances \label{tab3}}
\begin{tabular}{lc} \hline
Property                             &J2229$+$2725          \\ \hline
$T_{\rm e}$(O {\sc iii}), K          &   24800$\pm$900       \\
$T_{\rm e}$(O {\sc ii}), K           &   16000$\pm$400       \\
$T_{\rm e}$(S {\sc iii}), K          &   20800$\pm$700       \\
$N_{\rm e}$(S {\sc ii}), cm$^{-3}$    &1000$\pm$600         \\ \\
O$^+$/H$^+$$\times$10$^6$            &0.472$\pm$0.044 \\
O$^{2+}$/H$^+$$\times$10$^5$          &1.138$\pm$0.087 \\
O$^{3+}$/H$^+$$\times$10$^6$          &0.319$\pm$0.046 \\
O/H$\times$10$^5$                   &1.218$\pm$0.087 \\
12+log(O/H)                         &7.085$\pm$0.031     \\ \\
N$^{+}$/H$^+$$\times$10$^7$          &0.196$\pm$0.071 \\
ICF(N)                              &18.304 \\
N/H$\times$10$^7$                   &3.584$\pm$1.467 \\
log(N/O)                            &$-$1.535$\pm$0.180~ \\ \\
Ne$^{2+}$/H$^+$$\times$10$^6$        &1.805$\pm$0.130 \\
ICF(Ne)                             &1.029 \\
Ne/H$\times$10$^6$                  &1.856$\pm$0.139 \\
log(Ne/O)                           &$-$0.817$\pm$0.045~ \\ \\
S$^+$/H$^+$$\times$10$^7$            &0.129$\pm$0.013 \\
S$^{2+}$/H$^+$$\times$10$^7$          &0.768$\pm$0.260 \\
ICF(S)                              &3.844 \\
log(S/O)                            &$-$1.552$\pm$0.130~ \\ \\
Ar$^{2+}$/H$^+$$\times$10$^7$        &0.191$\pm$0.021 \\
ICF(Ar)                             &2.162 \\
Ar/H$\times$10$^7$                  &0.413$\pm$0.173 \\
log(Ar/O)                           &$-$2.469$\pm$0.185~ \\ \\
\hline
  \end{tabular}
  \end{table}

\section{Heavy element abundances}\label{sec:abundances}

We use here the emission-line fluxes derived from the LBT spectrum to derive heavy element abundances. We emphasize that the LBT spectrum is crucial for this task as some important lines, such as the [O~{\sc iii}] $\lambda$ 5007 and the H$\alpha$ lines, are clipped in the SDSS spectrum.
We follow the prescriptions of \citet{I06} to derive heavy element abundances in 
J2229$+$2725. The electron temperature $T_{\rm e}$(O~{\sc iii}) is calculated 
from the [O~{\sc iii}]$\lambda$4363/($\lambda$4959 + $\lambda$5007) 
emission-line flux ratio. It is used to derive the abundances of O$^{3+}$, 
O$^{2+}$ and Ne$^{2+}$.
The [S~{\sc iii}] $\lambda$9069 and [O~{\sc ii}] $\lambda$ 7320, 7330 emission
lines, which can, in principle, be used to directly derive the temperatures, are not detected
in the LBT spectrum of J2229+2725. Therefore, the abundances of O$^{+}$, N$^{+}$ and S$^{+}$ are derived with the electron
temperature $T_{\rm e}$(O~{\sc ii}) and the abundances of S$^{2+}$ and Ar$^{2+}$
are derived with the electron temperature $T_{\rm e}$(S~{\sc iii}), using the 
relations of \citet{I06} between $T_{\rm e}$(O~{\sc ii}), $T_{\rm e}$(S~{\sc iii})
and $T_{\rm e}$(O~{\sc iii}). The electron number density was derived from the
[S~{\sc ii}]$\lambda$6717/$\lambda$6731 flux ratio. The electron temperatures
and electron number densities are shown in Table~\ref{tab3}.
We note that the electron temperature $T_{\rm e}$(O~{\sc iii}) of 24,800 K 
is very high, implying that the oxygen
abundance is low. Furthermore, the number density of the ionised gas is also
very high, equal to 1000 cm$^{-3}$, which is atypical of SFGs, indicating a very young
age for the star-forming region.

The ionic abundances, ionisation correction factors and total O, N, Ne, S and 
Ar abundances are obtained using relations by \citet{I06}. They are 
presented in Table~\ref{tab3}. 

The nebular oxygen abundance of 12+logO/H = 7.085$\pm$0.031 in J2229$+$2725 is 
among the lowest found for SFGs, where O/H = O$^{+}$/H$^{+}$ + O$^{2+}$/H$^{+}$ 
+ O$^{3+}$/H$^{+}$. The N/O, Ne/O, S/O and Ar/O abundance ratios 
for this galaxy are similar to those in other 
low-metallicity SFGs \citep[e.g., ][]{I06}.


\section{Kinematics of the ionised gas}

The medium-resolution DIS spectrum allows us to put constraints on the 
kinematics of ionised gas in J2229$+$2725. In Fig.~\ref{fig3} is shown the
H$\alpha$ profile at observed wavelengths (black solid line). It is somewhat
broader than the instrumental profile shown by a black dotted line. 
Thus, the H$\alpha$ emission line is partially resolved. Its profile
is perfectly fitted by a single Gaussian profile (red solid line), indicating
a simple kinematical structure of the ionised gas with a regular motion, 
characteristic of a single H~{\sc ii} region without evident signs of merging 
or asymmetric infall or outflow.

\begin{figure}
\includegraphics[angle=-90,width=0.90\linewidth]{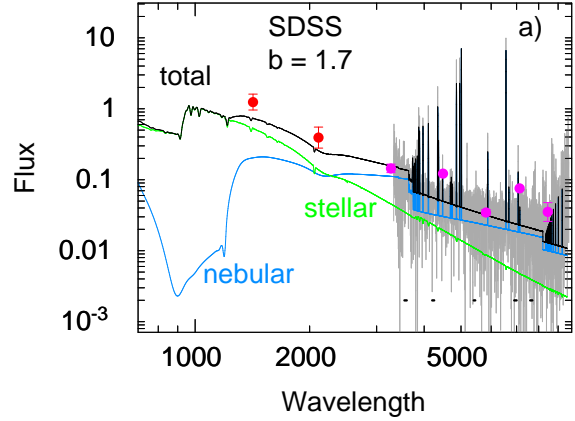}
\includegraphics[angle=-90,width=0.90\linewidth]{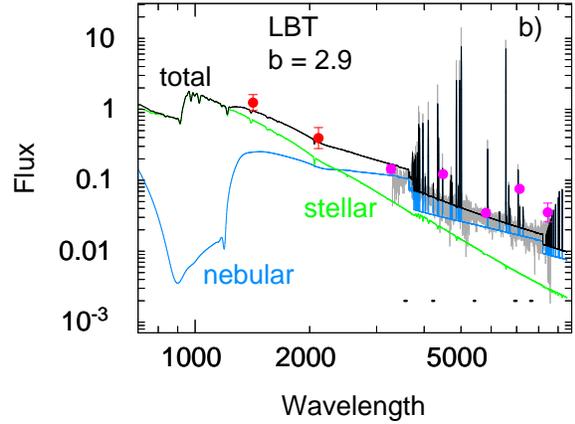}
\caption{{\bf a)} and {\bf b)} Best-fit spectral energy distributions 
obtained from fitting the rest-frame SDSS and LBT spectra, respectively, by 
randomly varying $t_{\rm b}$, $t_1$, $t_2$ and $b$~=~$M_{\rm old}$/$M_{\rm young}$, 
where $t_{\rm b}$ is the age of young stellar population, $t_2$ -- $t_1$ is the 
interval, in which the old stellar population was formed. 
In both panels, the observed spectra are shown in grey, 
the modelled stellar, nebular and total SEDs are shown by green, blue and black 
lines, respectively, and are labelled. {\sl GALEX} FUV, NUV and SDSS 
$u$, $g$, $r$, $i$ and $z$ photometric data, blueshifted adopting the J2229+2725 
redshift, are shown by red and magenta filled circles, respectively. Short 
horizontal lines represent intervals used for continuum fitting, which are free of 
emission lines. Fluxes are in units 10$^{-16}$ erg s$^{-1}$cm$^{-2}$\AA$^{-1}$ and
wavelengths are in \AA.}
\label{fig5}
\end{figure}

Using the relation
\begin{equation}
\sigma_{\rm int} = \frac{\sqrt{FWHM_{\rm obs}^2 - FWHM_{\rm inst}^2}}{2.355}
\frac{300000}{\lambda_{\rm obs}}\ \ {\rm km\ s}^{-1}
\label{sigint}
\end{equation}
we deconvolve the observed H$\alpha$ profile with the instrumental profile
and derive an intrinsic velocity dispersion $\sigma_{\rm int}$.
Here $FWHM_{\rm obs}$ and $FWHM_{\rm inst}$ are the measured full widths at half
maximum of the H$\alpha$ and instrumental profiles at observed wavelength
$\lambda_{\rm obs}$. We derive an intrinsic velocity dispersion 
$\sigma_{\rm int}$ = 34.7 km s$^{-1}$.
The intrinsic velocity dispersion consists of the thermal and turbulent
components. Adopting the electron temperature $T_{\rm e}$(O~{\sc iii}) = 24800~K
(Table~\ref{tab3}) and the relation
\begin{equation}
\sigma_{\rm th} = \sqrt{\frac{kT_{\rm e}({\rm O\ III})}{m}} \label{sigth}
\end{equation}
we obtain the thermal velocity dispersion for hydrogen atoms 
$\sigma_{\rm th}$ = 14.3 km s$^{-1}$. In Eq.~\ref{sigth}, $k$ is the Boltzman 
constant, $m$ is the proton mass. Finally, deconvolving the intrinsic
profile with a thermal profile
\begin{equation}
\sigma_{\rm tur} = \sqrt{\sigma_{\rm int}^2 - \sigma_{\rm th}^2}, \label{sigtur}
\end{equation}
we obtain the velocity dispersion for the turbulent component 
$\sigma_{\rm tur}$ = 31.6 km s$^{-1}$. 

Thus, the broadening of the H$\alpha$
emission line in J2229$+$2725 is dominated by macroscopic turbulent motions.

\section{Integrated characteristics of J2229$+$2725}
\label{sec:integr}

Everywhere in the paper, we have adopted the luminosity distance
$D_L$ = 347 Mpc, obtained from the galaxy redshift for the cosmological 
parameters H$_0$ = 67.1 km s$^{-1}$ Mpc$^{-1}$, $\Omega_m$ = 0.318, 
$\Omega_\Lambda$ = 0.682 \citep{P14}. It is used
to derive integrated characteristics such as stellar mass and luminosities.

The stellar mass of J2229$+$2725 is determined from fitting the 
spectral energy distribution (SED). 
We follow the prescriptions of \citet{I18a}, {\sc starburst99} models \citep{L99} 
and adopt stellar evolution models by \citet{G00},
stellar atmosphere models by \citet*{L97} and the \citet{S55} initial mass function (IMF) with lower and upper mass limits of 0.1 M$_\odot$ and 100 M$_\odot$,
respectively.

The equivalent width of the H$\beta$ emission line in J2229$+$2725 of 577\AA\
is exceptionally high (Table \ref{tab2}), indicating a very high contribution 
of the nebular continuum, amounting to more than 50 per cent of the total continuum
near the H$\beta$ emission line. Therefore, following e.g. \citet{I18a}, both 
the stellar and nebular continua are taken into account in the SED fitting.
The star-formation history in J2229$+$2725 was approximated 
by a recent short burst at age $t_{\rm b}$ $<$ 10 Myr and a prior 
continuous star formation with a constant SFR during the time interval
$t_2$ -- $t_1$ with $t_1$, $t_2$ $>$ 10 Myr and $t_2$ $>$ $t_1$. 
We use a Monte Carlo method with $\chi^2$ minimisation, varying $t_{\rm b}$,
$t_1$, $t_2$ and the mass fraction of stellar populations 
$b$ = $M_{\rm old}$/$M_{\rm young}$ formed after the age of 10 Myr and 
during the burst, and aiming to obtain the best agreement 
between the modelled and observed continuum over the entire wavelength range of 
the SDSS spectrum. Additionally, the observed EW(H$\beta$) and EW(H$\alpha$) 
have to be simultaneously reproduced by the best model.
More details on the SED fitting procedure can be found e.g. in \citet{I18a}.

We use both the continua of SDSS and LBT spectra, avoiding regions with
emission lines, to fit SEDs and to derive stellar masses. 
The best fits are shown in Fig.~\ref{fig5}a for the SDSS spectrum and 
in Fig.~\ref{fig5}b for the LBT spectrum. It is clear that the LBT spectrum has a considerably better signal to noise ratio than the SDSS spectrum. Thus, the physical parameters derived from the LBT data are more reliable, and we shall adopt them throughout the paper. However, we discuss also in this section the parameters derived from the SDSS spectrum for comparison. 

The derived parameters $t_{\rm b}$,
$t_1$, $t_2$, $b$ and stellar masses $M_{\rm young}$, $M_{\rm old}$ and 
$M_\star$~=~$M_{\rm young}$~+~$M_{\rm old}$ of the SED fits are presented in 
Table~\ref{tab4} (second column is for the SDSS spectrum and fifth column is for
the LBT spectrum). We note that the photometric SDSS and {\sl GALEX} data are not
used for SED fitting. The SDSS photometric fluxes (magenta symbols in 
Fig.~\ref{fig5}) are derived from model magnitudes extracted from the SDSS 
data base. The model magnitudes are calculated using the best-fit parameters 
of the galaxy brightness profile in 
the $r$ band, and applying them to all other bands; the light is therefore 
measured consistently through the same aperture in all bands.
These magnitudes are very similar to the PSF magnitudes presented in
the same data base. The SDSS photometric fluxes in $g$, $i$ and $z$ bands 
are considerably higher than the 
fluxes in the continuum at the same wavelengths, due to the contribution of strong 
emission lines, whereas the contribution of emission lines in the $u$ and $r$ bands
is small. The fluxes in {\sl GALEX} FUV and NUV photometric bands can be used 
to check the quality of SED fitting in the UV range. It is seen in 
Fig.~\ref{fig5}a that observed FUV and NUV fluxes (red symbols) are slightly 
above the fluxes modelled with the SDSS spectrum (black line), whereas 
the agreement is much better for the fit to the LBT spectrum (Fig.~\ref{fig5}b).

The remarkable feature of J2229$+$2725 is that the continuum in the visible
range is dominated by nebular emission (blue line in Fig.~\ref{fig5}), which is 
several times higher than the stellar emission (green line in 
Fig.~\ref{fig5}). Its 
contribution to the total emission in the continuum increases with increasing  
wavelength. Nebular emission in this galaxy is considerable even in the UV 
range longward of the Ly$\alpha$ line, due to sharply rising two-photon emission
at $\lambda$~$\geq$~1215.67~\AA. It should be taken into account e.g. in
the determination of the UV stellar slope $\beta$, commonly used in studies of 
star-forming galaxies.

\begin{table*}
\caption{Results of SED fitting \label{tab4}}
\begin{tabular}{lcccccc} \hline
\multicolumn{1}{c}{Parameter}&  \multicolumn{3}{c}{SDSS} & \multicolumn{3}{c}{LBT}       \\
\multicolumn{1}{c}{(1)}&(2)&(3)&(4)&(5)&(6)&(7) \\
 \hline
$b$ = $M_{\rm old}$/$M_{\rm young}$&~\,1.7$^{\ddag}$&~~\,0.0$^{\ddag\ddag}$&~~\,8.0$^{\ddag\ddag}$&
~\,2.1$^{\ddag}$&~~\,0.0$^{\ddag\ddag}$&~~\,8.0$^{\ddag\ddag}$ \\
$t_{\rm b}$, Myr$^{*}$    &  0.4 & 2.0&0.7& 0.4& 2.1&0.7 \\
$t_1$, Gyr$^{**}$        &  0.8  & ...&9.7& 1.4& ...&9.7 \\
$t_2$, Gyr$^{**}$        &  2.8  & ...&12.5& 5.5& ...&12.5 \\
log $M_{\rm young}$/M$_\odot$$^{\dag}$&   6.42 & 6.35&6.40& 6.37&6.29&6.33 \\
log $M_{\rm old}$/M$_\odot$$^{\dag\dag}$ &   6.60 & ... &7.30& 6.83& ...&7.24 \\
log $M_\star$/M$_\odot$$^{\dag\dag\dag}$&   6.82&6.35&7.36&6.96&6.29&7.29 \\
\hline
  \end{tabular}



\hbox{$^{\ddag}$Best-fit value of the old-to-young stellar population
mass ratio.}

\hbox{$^{\ddag\ddag}$Fixed value of $b$. The value $b$ = 0.0 corresponds to the
model with zero mass of the old stellar population. On the other hand,}

\hbox{the 
value $b$ = 8.0 corresponds to the model in which the old stellar 
population consists only of the least luminous stars}

\hbox{in the age range between $t_1$ $\sim$ 10 Gyr and $t_2$ $\sim$ 13 Gyr.
}

\hbox{$^{*}$$t_{\rm b}$ is the age of young stellar population.}

\hbox{$^{**}$$t_2$ -- $t_1$ is the interval, in which the stellar population is
formed.}

\hbox{$^{\dag}$$M_{\rm young}$ is the mass of young stellar population.}

\hbox{$^{\dag\dag}$$M_{\rm old}$ is the mass of old stellar population.}

\hbox{$^{\dag\dag\dag}$$M_\star$ is the total stellar mass derived from the 
extinction-corrected spectrum.}

  \end{table*}

The mass of the young stellar population $M_{\rm young}$ of 
$\sim$~10$^{6.4}$~M$_\odot$ is consistently derived from both the SDSS and LBT 
spectra (Table~\ref{tab4}), despite the fact that the SDSS spectrum is considerably 
noisier (compare Fig.~\ref{fig5}a and \ref{fig5}b). The determination of the
mass of old stellar population $M_{\rm old}$ is much more uncertain because of
its negligible contribution to emission in the visible range. We obtain a
stellar mass for J2229$+$2725 as low as $M_\star$ = 10$^{6.96}$~M$_\odot$ from the
best-fit SED to the LBT spectrum (Table~\ref{tab4}). 
This yields a very high specific star formation rate sSFR of 
$\sim$ 75 Gyr$^{-1}$, indicative of very active ongoing star formation.

For the sake of
comparison, we also derive best-fit SEDs adopting $b$ = 0, corresponding to
$M_{\rm old}$ = 0, with parameters presented in the third and sixth columns of 
Table~\ref{tab4}. In this case we derive the lower stellar mass limit 
$M_\star$ = $M_{\rm young}$ $\sim$ 10$^{6.3}$ $M_\odot$.

On the other hand, 
assuming that the old stellar population consists only of the least
luminous stars in the age range of $\sim$ 10 -- 13 Gyr, we can set an
upper limit for the total stellar mass $M_\star$. For J2229$+$2725 it can not be 
higher than $\sim$ 10$^{7.3}$~M$_\odot$, corresponding to $b$ $\sim$ 8
(see the other parameters for the best-fit models with $b$ = 8 in 
columns four and seven of Table~\ref{tab4}). Otherwise, 
the high observed equivalent widths of the H$\beta$ and H$\alpha$ emission lines
can not be reproduced with our fitting procedure because of a too high 
contribution to the continuum by the emission of the old stellar population.

We note that, despite a wide range of $b$ from 0 to 8 in the considered models, their
best-fit SEDs are nearly indistinguishable, with only slightly different $\chi^2$.
The adopted models with randomly varied $b$'s have the lowest values of 
$\chi^2$. Their SEDs are shown in Fig.~\ref{fig5}.

The absolute SDSS $g$ magnitude, corrected for the Milky Way 
extinction is $M_g$ = $-$16.39 mag (Table \ref{tab1}). Thus, J2229$+$2725
is the second most luminous compact SFG known with 12~+~logO/H $<$ 7.1,
after J1234$+$3901.

\begin{figure}
\centering{
\includegraphics[angle=-90,width=0.90\linewidth]{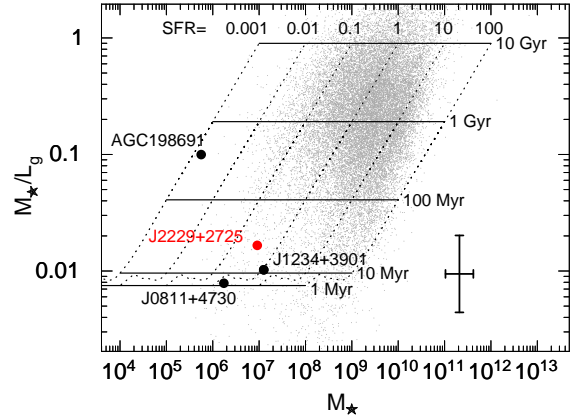}
}
\caption{Relation between the stellar mass-to-luminosity ratio and the stellar
mass of SFGs. All quantities are expressed in solar units. Selected 
lowest-metallicity SFGs J0811+4730, J1234+3901 and AGC~198691 with 
12+logO/H $\sim$ 7.0 are represented by labelled black filled circles. 
The galaxy J2229+2725 is shown by a red filled circle. The error bars indicate 
the errors of $M_\star$ and $M_\star$/$L_g$ for that SFG. For comparison, compact 
SFGs from the SDSS DR16 with redshifts $z$ $>$ 0.01 (grey dots) are also shown. 
Dotted lines 
represent Starburst99 models with continuous star formation and a SFR varying 
from 0.001 to 100 M$_\odot$ yr$^{-1}$, during periods from the present to 10 
Gyr, 1 Gyr, 100 Myr, 10 Myr, and 1 Myr in the past (horizontal solid 
lines). These models include the contribution of both the stellar and nebular 
continua.}
\label{fig6}
\end{figure}

Adopting $M_\star$ = 10$^{6.96}$~M$_\odot$, we derive the very low 
mass-to-luminosity ratio $M_\star$/$L_g$ of 0.0166
(in solar units) for J2229$+$2725. It is comparable to those of two other most
metal-poor compact SFGs, J0811$+$4730 and J1234$+$3901 \citep{I18a,Iz19a}. 
However, it is approximately one order of magnitude lower than the 
mass-to-luminosity 
ratio of another extremely metal-poor galaxy SFG AGC~198691 \citep{H16}, 
which contains a higher fraction of older stellar population (between 100 Myr 
and 1 Gyr), and up to two orders of magnitude lower than the mass-to-light 
ratios of the majority of compact SFGs in the SDSS which contain even more 
older stellar populations with ages between 100 Myr and 10 Gyr (see Fig.~\ref{fig6}).
Such a low value is consistent with the mass-to-luminosity ratio of a zero-age 
instantaneous burst, meaning that the contribution of older populations
to the galaxy's luminosity is negligible compared to the starburst luminosity, 
at least, for the emission inside the spectroscopic aperture,
in agreement with the galaxy's extremely high EW(H$\beta$). 

\begin{figure}
\centering{
\includegraphics[angle=-90,width=0.90\linewidth]{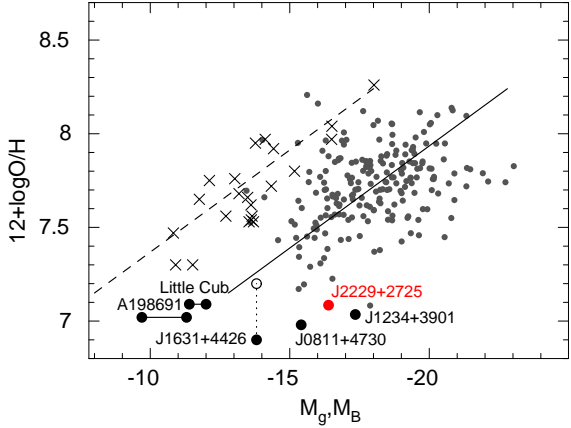}
}
\caption{The oxygen abundance - absolute magnitude diagram. 
The most metal-poor galaxies (the same ones as in Fig.~\ref{fig4}) are represented by filled 
and open circles. Low-luminosity 
galaxies are represented by crosses, with the fit to these data by \citet{B12} shown by a dashed line. 
The solid line is the fit to the SDSS compact SFGs from DR16 with 
EW(H$\beta$)~$\geq$~200\AA\ and oxygen 
abundances derived by the direct $T_{\rm e}$ method (grey dots). This subsample
of SDSS galaxies includes only compact galaxies with
[O~{\sc iii}]$\lambda$4363 emission lines in their spectra, measured with 
errors less than 20 per cent of their fluxes. The vertical dotted line 
for J1631$+$4426
connects the locations of the galaxy with oxygen abundances derived by the 
direct $T_{\rm e}$ method (filled circle) and by the strong-line method (open circle,
see Sect.~\ref{SLM}). The horizontal lines connect the positions of the nearby galaxies 
AGC~198691 and Little Cub, calculated with two different distances.}
\label{fig7}
\end{figure}

\section{Luminosity -- metallicity relation}\label{lummet}

It has been established  \citep[e. g. ][and references therein]{I18a}
that extremely metal-deficient galaxies with very active star formation 
deviate from the luminosity -- metallicity relation defined by the 
bulk of the SDSS SFGs. The XMD galaxy J2229$+$2725 shows the same behavior, being too luminous for its oxygen abundance 
(Fig.~\ref{fig7}). Oxygen abundances for all galaxies shown in Fig.~\ref{fig7} 
have been derived by the direct $T_{\rm e}$ method. In particular, only compact 
SFGs from the SDSS DR16 (grey dots) with an [O~{\sc iii}]$\lambda$4363 emission line 
detected in their spectra with an accuracy better than 20 per cent are
included. The two galaxies with the lowest luminosities, Little Cub 
\citep{H17} and AGC~198691 \citep{H16}, deviate slightly from the relation by
\citet{B12} for relatively quiescent SFGs (crosses and dashed line in 
Fig.~\ref{fig7}). On the other hand, all other lowest-metallicity SFGs, with higher luminosities, 
including J2229$+$2725, strongly deviate from the \citet{B12} relation. 

It has been suggested e.g. by \citet*{I12} and \citet{G17} 
that these deviations can be explained by the enhanced brightnesses of the 
galaxies undergoing active star formation and they are higher for galaxies
with high EW(H$\beta$) and correspondingly with younger starburst age.
In particular, \citet{I18a} have shown that the mass-to-luminosity ratio
of the most extreme galaxies, like J2229$+$2725, are lower by a factor of more than 10 
than typical ratios of SDSS galaxies. This difference translates to at least
$\sim$ 2 -- 3 brighter magnitudes for the XMD starbursts
and it can explain the offset of these galaxies from the relation for the bulk of
compact SFGs from the SDSS (solid line in Fig.~\ref{fig7}). On the other hand,
SDSS compact galaxies are offset by $\sim$ 3 mags to brighter magnitudes from 
the sample of quiescent dwarf galaxies studied by \citet{B12}, indicating that the 
luminosity of the latter galaxies is dominated not by young stellar populations in starburst regions, but by older stellar populations.

It has also been suggested that XMD galaxies 
have too low metallicities for their high luminosities. These low metallicities may have been caused by the infall of metal-poor gas from galactic halos \citep{EC10}.

\section{Modified strong-line method for the determination of oxygen 
abundance}\label{SLM}

We have pointed out that J2229$+$2725 possesses the extraordinarily high 
O$_{32}$ ratio of 53. We show in this section how such an object can be used to 
improve substantially the calibration of the strong-line method to derive 
oxygen abundances in star-forming galaxies, including those with extreme
excitation conditions.

\begin{figure*}
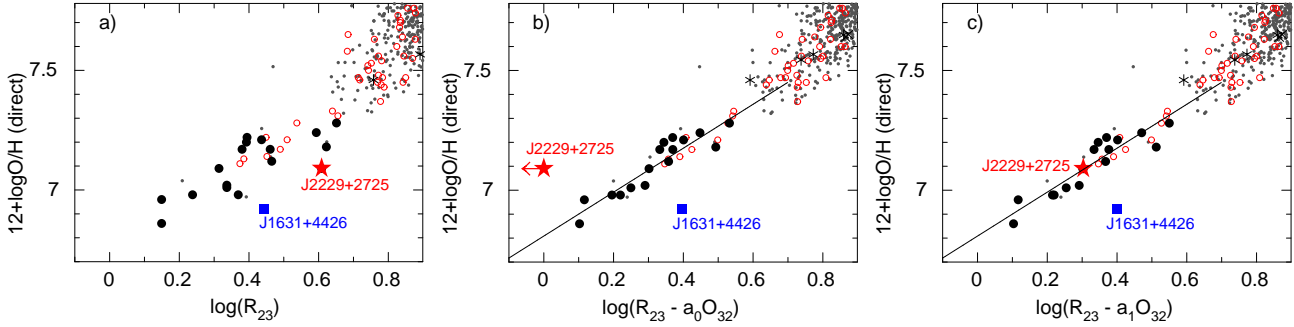

\hbox{
\includegraphics[angle=-90,width=0.32\linewidth]{R23_O32_DR14log_1.ps}
\includegraphics[angle=-90,width=0.32\linewidth]{R23_O32_O_DR14log_0.ps}
\includegraphics[angle=-90,width=0.32\linewidth]{R23_O32_O_DR14log_1.ps}
}
\caption{The {\bf a)} (log R$_{23}$) - (12 + logO/H), {\bf b)} 
log(R$_{23}$ - $a_0$O$_{32}$) - (12 + logO/H), and {\bf c)} 
log(R$_{23}$ - $a_1$O$_{32}$) - (12 + logO/H) relations, where $a_0$ = 0.080
\citep{Iz19b} and $a_1$ = 0.080 -- 0.00078O$_{32}$ (this paper). In all these 
relations 12 + logO/H is derived by the direct method, R$_{23}$ = 
([O~{\sc iii}]$\lambda$4959+$\lambda$5007 + [O~{\sc ii}]$\lambda$3727)/H$\beta$ and 
O$_{32}$ = [O~{\sc iii}]$\lambda$5007/[O~{\sc ii}]$\lambda$3727. The galaxy 
J2229$+$2725 is shown 
by a red filled star, while J1631$+$4426 \citep{Ko20} is represented by a blue 
filled square. Other lowest-metallicity SFGs from \citet{I18a}, the galaxy 
Little Cub \citep{H17}, the galaxy J1234$+$3901 from \citet{Iz19a} and the 
H~{\sc ii} region DDO~68\#7 \citep{An19}
are shown by filled circles. SFGs with the highest O$_{32}$ ratios in the 
range $\sim$ 20 - 40 \citep{I17} are shown by asterisks. Samples of SFGs used 
for the He abundance determination \citep[][and references therein]{I14} are 
shown by open circles. Grey dots represent SFGs at
$z$ $>$ 0.02 from the SDSS, with the [O~{\sc iii}]$\lambda$4363 emission line 
measured with an accuracy better than 20 per cent. The solid lines in {\bf b)} 
and {\bf c)} are linear fits to the data with 
log(R$_{23}$ -- $a_0$O$_{32}$) $\leq$ 0.7 and 
log(R$_{23}$ -- $a_1$O$_{32}$) $\leq$ 0.7, respectively, and excluding 
J1631$+$4426.}
\label{fig8}
\end{figure*}

The most reliable method for oxygen abundance determination is the direct 
method, but it requires the detection, with good accuracy, of the 
[O~{\sc iii}]$\lambda$4363 emission line for the electron temperature determination. 
However, most low-metallicity SFGs are intrinsically faint objects
with a weak or undetected [O~{\sc iii}]$\lambda$4363 emission line, precluding a 
reliable oxygen abundance determination by the direct method.
Many authors, including most recently e.g. \citet{Iz19b}, have developed a so-called "strong-line method" with the use of strong
oxygen lines, which is calibrated using high-quality observations of the most
metal-poor galaxies with a well-detected [O~{\sc iii}]$\lambda$4363 emission line.

A common problem of the strong-line method is that it depends on several
parameters. In particular, the method based on oxygen line intensities 
depends not only on the sum R$_{32}$ = $I$([O~{\sc ii}]$\lambda$3727 + 
[O~{\sc iii}]$\lambda$4959 + [O~{\sc iii}]$\lambda$5007])/$I$(H$\beta$), but also on the 
ionisation parameter. A good indicator of the ionisation parameter is the
O$_{32}$ ratio. The dependences of 12 + logO/H on various combinations of 
oxygen line intensities are illustrated in Fig.~\ref{fig8}, where we basically
show the data of \citet{Iz19b}, adding the two most recently discovered XMD galaxies,
J1631$+$4426 by \citet{Ko20} and J2229$+$2725 discussed here. The oxygen abundances of all
galaxies in Fig.~\ref{fig8} are derived by the direct method and the 
[O~{\sc iii}]$\lambda$4363 emission line is measured with errors less than 20 per cent
in their fluxes.
The log R$_{23}$ -- 12~+~logO/H diagram is shown in Fig.~\ref{fig8}a. It is 
important that the data include objects with the largest O$_{32}$ ratios, 
indicators of a very high ionisation parameter. The use of these galaxies 
allows us to better eliminate the dependence of the calibration relation on
the ionisation parameter. The dispersion of data at the lowest-metallicity end
of the diagram, with 12 + logO/H $\la$ 7.3, 
is relatively high and it is caused by different ionisation
parameters in the galaxies. The most deviant objects are J2229$+$2725, with the 
highest O$_{32}$, and J1631$+$4426.

To minimise the data scatter, \citet{Iz19b} have proposed to use instead the diagram 
log(R$_{23}$ -- $a_0$O$_{32}$) -- (12~+~logO/H), with $a_0$ = 0.080 (Fig.~\ref{fig8}b). The data in Fig.~\ref{fig8}b, excluding the two deviant points J2229$+$2725
and J1631$+$4426, can be well fitted by the linear relation
\begin{equation}
12+\log\frac{\rm O}{\rm H} = 0.950 \log({\rm R}_{23} - a_0{\rm O}_{32}) + 6.805.
\label{eq1}
\end{equation}
shown by a solid line.

It is seen in Fig.~\ref{fig8}b that the majority of the extremely low-metallicity SFGs
follow tightly the relation Eq.~\ref{eq1}. The scatter is small, much
lower than that in Fig.~\ref{fig8}a. Exceptions are the two SFGs mentioned above. One of them, 
J1631$+$4426, has a low O$_{32}$ = 3.41 \citep{Ko20}, therefore its 
ionisation parameter correction is small and will not change much its location in Fig.~\ref{fig8}b, compared to that in Fig.~\ref{fig8}a. 
On the other hand, the ionisation parameter correction, based on 
Eq.~\ref{eq1}, increases with increasing O$_{32}$, moving the galaxy to 
the left. Thus, for the SFG J2229$+$2725 with the highest 
O$_{32}$ = 53, we find that (R$_{23}$ -- $a_0$O$_{32}$) takes on a 
negative value,
making Eq.~\ref{eq1} not applicable. The location of J2229$+$2725 shown in 
Fig.~\ref{fig8}b corresponds to a slightly lower O$_{32}$ of 50.7, at which 
(R$_{23}$ -- $a_0$O$_{32}$) is still positive. 

The large deviation of 
J2229$+$2725 from other galaxies thus suggests that Eq.~\ref{eq1} needs to be 
modified. The modification is found by minimising 
the scatter of galaxies, including J2229$+$2725. We found the relation
\begin{equation}
12+\log\frac{\rm O}{\rm H} = 0.950 \log({\rm R}_{23} - a_1{\rm O}_{32}) + 6.805,
\label{eq2}
\end{equation}
where $a_1$ = 0.080 -- 0.00078O$_{32}$. This relation is shown in 
Fig.~\ref{fig8}c. It is seen that the SFG J2229$+$2725 follows very closely this
relation. 

On the other hand, the galaxy J1631$+$4426 remains an outlier.
\citet{Ko20} have derived 12 + logO/H = 6.90 $\pm$ 0.03 for this galaxy, making it the most metal-poor SFG known. However, this very low oxygen abundance is 
not supported by Eq.~\ref{eq2}. 
Using this equation we derive 
12 + logO/H = 7.18 for J1631$+$4426, in full agreement with the value derived
by \citet{Ko20} with the use of strong-line calibration. If so, J1631$+$4426
can not be considered as the most metal-deficient nearby SFG known. However, new
high quality observations of J1631$+$4426 are needed to confirm this conclusion.

Eq.~\ref{eq2} can be used to search for extremely 
low-metallicity galaxies with oxygen abundances extending to values 12~+~logO/H~$<$~7.0, a range where other
methods are not applicable because of the weakness of the emission lines. In addition, 
it can be used to verify the abundances derived by the direct method.

\section{Conclusions}\label{sec:conclusions}

We present Large Binocular Telescope (LBT)/Multi-Object Dual Spectrograph (MODS)
spectrophotometric observations of the compact star-forming galaxy (SFG)
J2229$+$2725, selected from the Data Release 16 (DR16) of the Sloan Digital Sky
Survey (SDSS). This local SFG possesses extraordinary properties 
(in particular a very low metallicity and an extremely high O$_{32}$ ratio), 
which makes it stand apart from other galaxies. LBT observations have been
supplemented by medium-resolution spectroscopy with the Dual Imaging 
Spectrograph (DIS) mounted on the 3.5 meter telescope of the Apache Point 
Observatory (APO). Our main results are as follows.

1. The emission-line spectrum of J2229$+$2725, with a stellar mass of 
9.1$\times$10$^6$ M$_\odot$ and an absolute SDSS $g$-band magnitude 
of --16.39 mag, originates from a very dense ($N_{\rm e}$(S~{\sc ii}) =
1000~cm$^{-3}$) and hot ($T_{\rm e}$(O~{\sc iii}) = 24800~K) ionised gas.
The detection of the He~{\sc ii} $\lambda$4686 emission line with the flux
ratio $I$(He~{\sc ii} $\lambda$4686)/$I$(H$\beta$) of $\sim$~0.021 indicates
the presence of hard ionising radiation.
The oxygen abundance in J2229$+$2725 is 12+logO/H = 7.085$\pm$0.031, 
one of the lowest for nearby SFGs. 

2. The rest-frame equivalent width EW(H$\beta$) of the H$\beta$ 
emission line of 577\AA\ in the J2229$+$2725 spectrum is the highest measured
so far in SFGs. The mass-to-luminosity ratio $M_\star$/$L_g$ in J2229$+$2725 of 
$\sim$~0.017 in solar units is extremely low.
Both the extremely high EW(H$\beta$) and extremely low $M_\star$/$L_g$
indicate that the J2229$+$2725 emission is strongly dominated by a very young 
stellar population with an age $\la$ 2 Myr. These properties are 
similar to those of the two most metal-deficient SFGs known, J0811$+$4730 and 
J1234$+$3901 \citep{I18a,Iz19b}. 

3. The O$_{32}$ = $I$([O~{\sc iii}]$\lambda$5007)/$I$([O~{\sc ii}]$\lambda$3727)
flux ratio of $\sim$ 53 in J2229$+$2725 is extremely high. It is the highest 
ever measured for low-metallicity SFGs. These properties together with
the extremely high rest-frame equivalent width EW(H$\beta$) of the H$\beta$
emission line and the very dense H~{\sc ii} region imply that this galaxy 
might be at a very young stage of its formation. If so, then J2229$+$2725
is one of the best local counterparts of young high-redshift dwarf SFGs.

4. J2229$+$2725 strongly deviates from the
luminosity-metallicity relation defined by the bulk of compact SDSS SFGs.
It is $\sim$ 2.5 times more metal-poor for its SDSS $g$-band absolute magnitude.

5. A medium resolution spectrum of J2229$+$2725 obtained with the Double 
Imaging Spectrograph mounted on the 3.5 meter Apache Point Observatory (APO) 
telescope shows that the ionised gas kinematics are dominated by macroscopic 
turbulent motions with a velocity dispersion of 31.6 km s$^{-1}$, with no 
evident sign of merging or asymmetric infall or outflow. 

6. The extremely high O$_{32}$ ratio in J2229$+$2725 allowed us to use the 
latter to improve the calibration of the strong-line method of oxygen abundance 
determination in extremely low-metallicity SFGs (with 12 + logO/H $<$ 7.45), 
which is based on the fluxes of 
strong nebular emission lines [O~{\sc ii}]$\lambda$3727 and 
[O~{\sc iii}]$\lambda$4959+$\lambda$5007 emission lines. 

7. The proposed strong-line method can be used to search for SFGs at the 
extreme end of oxygen abundances 12 + logO/H~$<$~7.0, where other methods, 
including the direct $T_{\rm e}$ method, can not be applied because of the 
weakness of emission lines. This method can also
be applied to verify oxygen abundances derived by the direct $T_{\rm e}$ method.
In particular, using our new calibration of the strong-line method, we find
that the oxygen abundance for the galaxy J1631$+$4426 is 0.28 dex higher than 
the value 12~+~logO/H = 6.90$\pm$0.03 derived by \citet{Ko20} using the direct $T_{\rm e}$ 
method, but it is consistent with their value derived using the strong-line
method.

\section*{Acknowledgements}

We are grateful to the anonymous referee for useful comments on the manuscript.
Y.I.I. and N.G.G. acknowledge support from the National Academy of Sciences of 
Ukraine by its priority project No. 0120U100935 ``Fundamental properties of 
the matter in the relativistic collisions of nuclei and in the early Universe''.
Funding for the Sloan Digital Sky Survey IV has been provided by
the Alfred P. Sloan Foundation, the U.S. Department of Energy Office of
Science, and the Participating Institutions. SDSS-IV acknowledges
support and resources from the Center for High-Performance Computing at
the University of Utah. The SDSS web site is www.sdss.org.
SDSS-IV is managed by the Astrophysical Research Consortium for the 
Participating Institutions of the SDSS Collaboration. 
This research has made use of the NASA/IPAC Extragalactic Database (NED), which 
is operated by the Jet Propulsion Laboratory, California Institute of 
Technology, under contract with the National Aeronautics and Space 
Administration.

\section*{Data availability}

The data underlying this article will be shared on reasonable request to the 
corresponding author.

\bsp

\label{lastpage}

\end{document}